\pdfoutput=1 
\documentclass{JINST}

\title{Study of RPC bakelite electrodes and detector performance for INO-ICAL}

\author{Ashok Kumar\thanks{Corresponding author.}~, Ankit Gaur, Md. Hasbuddin, Praveen Kumar, Purnendu Kumar, Daljeet Kaur, Swati Mishra and Md. Naimuddin\\
\llap \noindent Department of Physics and Astrophysics\\
University of Delhi, Delhi, India - 110007\\
E-mail: \email{ashok.hep@gmail.com}}

\abstract{The Resistive Plate Chambers (RPCs) are going to be used as the active detectors in the India-based Neutrino Observatory (INO)-Iron Calorimeter (ICAL) experiment for the detection and  study of atmospheric neutrinos. In this paper, an extensive study of structural and electrical properties  for different kind of bakelite RPC electrodes is presented. RPCs fabricated from these electrodes are tested for their detector efficiency and noise rate. The study concludes with the variation of efficiency, leakage current and counting rate over the period of operation with different gas compositions and operational conditions like temperature and relative humidity.}


\begin{document}

\section{Introduction}\label{sec:xxx}
The Resistive Plate Chambers (RPCs) \cite{santonico} have been used in various high energy physics experiments because of their excellent time resolution and tracking capabilities. The glass based RPCs are going to be used in the first phase of India-based Neutrino Observatory (INO)-Iron Calorimeter (ICAL) experiment as active detector elements for the detection and detail study of atmospheric neutrinos \cite{pramana} \cite{atmnu}. The R$\&$D on bakelite based RPCs is being explored for the future phases of ICAL \cite{meghna}. For a better and optimized performance of RPCs detectors, there is a need for exploring various aspects related to the nature of RPC electrode material and its dependence upon operating conditions. The single gap bakelite RPCs have been fabricated with samples procured locally from the available bakelite qualities, namely,  Formica and Hylam. Before fabrication of the RPC detector, the structural and electrical properties are studied for better understanding of the nature of material. Finally, RPCs are tested with cosmic hodoscope for their efficiency and rate measurements. 				 

\section{Electrode material characterization}\label{electrode}
The selection of appropriate RPC electrode, based upon the surface quality of electrodes, plays a significant role in the detector performance. The surface smoothness is crucial in reducing spontaneous discharge, which might affect the rate capability and various parameters related to the detector. It is established that the rough surface of electrode is a source for causing field emission which results in high leakage current \cite{dc1,dc2}. The non-uniformity, even for short ranges, is prone to distortion of the uniform electric field. Also, long range surface defects may often results in discharges that damage and degrade the RPC performance. 

The X-ray Diffractometry (XRD) spectrum gives hump like structure for both the samples. Hence, the bakelite samples are clearly amorphous in nature. There is not much difference found  in the elemental compositions (as shown in Table~\ref{tab:EDS}) through Energy Dispersive X-ray Spectroscopy (EDX) study for both the bakelite samples, except for the presence of small amounts of Al and Fe in Hylam. 

\begin{table} [ht] 
\centering 
\begin{tabular}{|c|c|c|c|}
\hline
Serial No. & Name of element & \multicolumn{2}{|c|}{Composition (percentage)}   \\ 
\hline 
         &          & Formica & Hylam    \\

\hline 
 1 &     C &33.22	 & 32.77	\\
 2 &     Na & 0.04 &	0.09 \\
  3&     Al & -	& 0.25	\\
  4&     Si & 0.02	& 0.18	\\
  5&     Ca & 0.05	& -	\\
  6&     O  & 66.57	& 66.5	\\
  7&     F & 0.07	& -	\\
  8&     S & 0.01	& 0.03	\\
  9&     Cl & 0.02	& -	\\
  10&     Fe & -	& 0.17	\\
\hline   

\end{tabular} 
\caption{The percentage composition of elements derived using EDX in both the bakelite samples.}
\label{tab:EDS}  
\end{table}

\begin{figure}[tbp]
\centering
\resizebox{7.0cm}{6.0cm}{\includegraphics{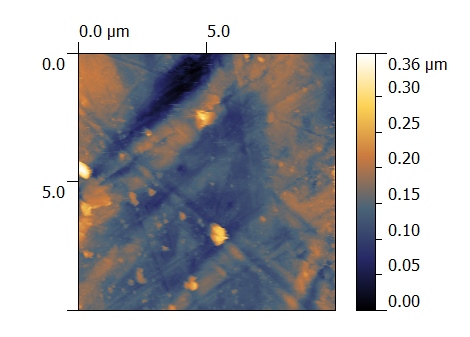}}
\resizebox{6.0cm}{5.0cm}{\includegraphics{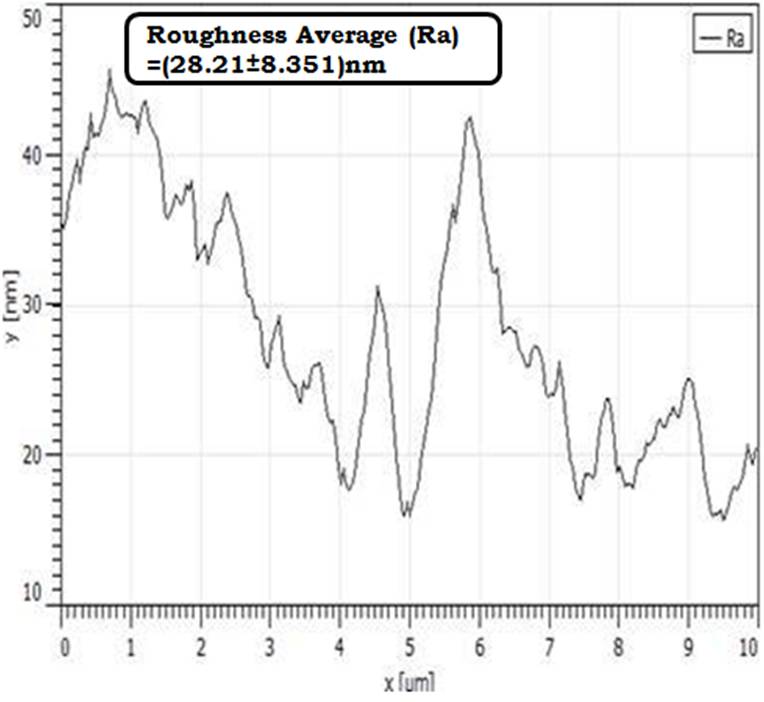}}
\scriptsize{\caption{\label{wocoat}Results of AFM study for Formica bakelite without silicon coating.}}
\end{figure}

\begin{figure}[tbp]
\centering
\resizebox{7.0cm}{6.0cm}{\includegraphics{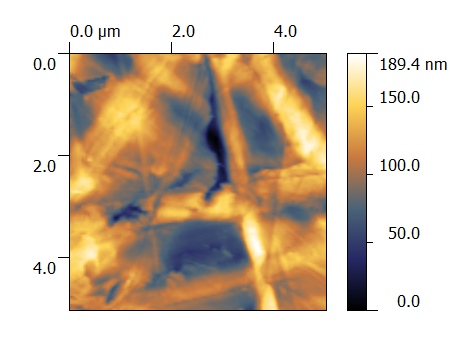}}
\resizebox{6.0cm}{5.0cm}{\includegraphics{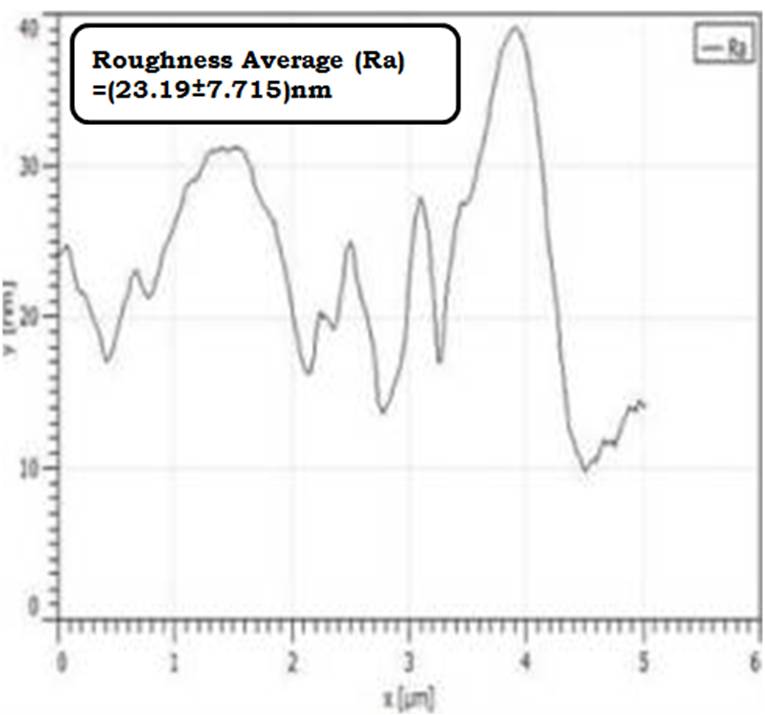}}
\scriptsize{\caption{\label{wcoat} Results of AFM study for Formica bakelite with silicon coating. The silicon coating on bakelite decreases the surface roughness from 28 nm to 23 nm.}}
\end{figure}

For the measurement of surface uniformity of each bakelite, various characterization analysis, namely, Scanning Electron Microscopy (SEM) and Atomic Force Microscopy (AFM) have been performed. AFM study has been done separately for Formica bakelite without silicon coating as shown in  Figure~\ref{wocoat}, and with silicon coating as shown in Figure~\ref{wcoat}. From AFM analysis, it is concluded that silicon coating helps in improving surface smoothness significantly.

\section{Resistivity measurement}\label{morphology}
The bakelite sheets are phenolic resin bonded malamine laminates. The resistivity plays an important parameter \cite{imppar} in the detector operation. The bulk and surface resistivities are measured using different electrical circuits. The bulk resistivity is obtained by measuring the leakage current flowing through a 1 cm$^3$ of insulating material and expressed in $\Omega$ cm whereas surface resistivity is obtained by measuring the electrical resistance of the surface of a  material and expressed in $\Omega/\square$ .

\subsection{\it Bulk resistivity}\label{bulk} 
\begin{figure}
\centering
\includegraphics[height=5cm,width=5cm]{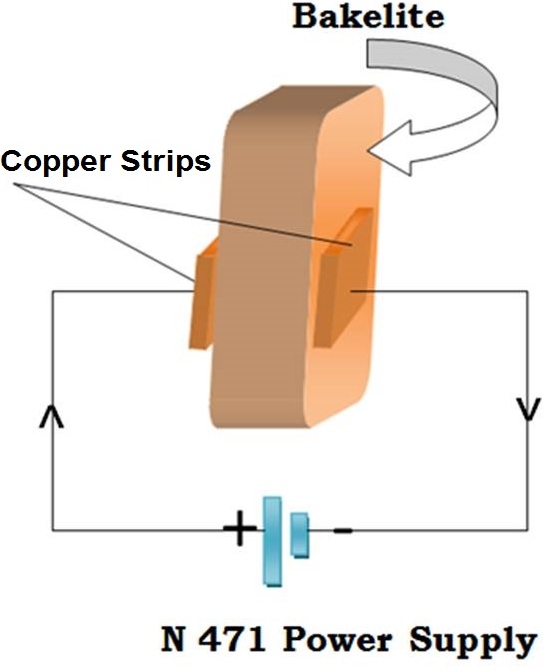}
\caption{\label{bset} The set-up for bulk resistivity measurement.}
\end{figure}

High bulk resistivity is an important parameter in the proper choice of the RPC electrode material for controlling the time resolution, counting rate, and leakage current \cite{meghna} of the detector. The bulk resistivity of the bakelite sample is measured using the set-up shown in  Figure~\ref{bset}. The bakelite sample is placed between two copper electrodes and the potential is applied using CAEN N471 power supply. The current distributed through the volume of the sample is then measured. The bulk resistivity is determined by the straight line fit to the current-voltage (I-V) curve. 

\begin{figure}
\centering
\includegraphics[height=7cm,width=9cm]{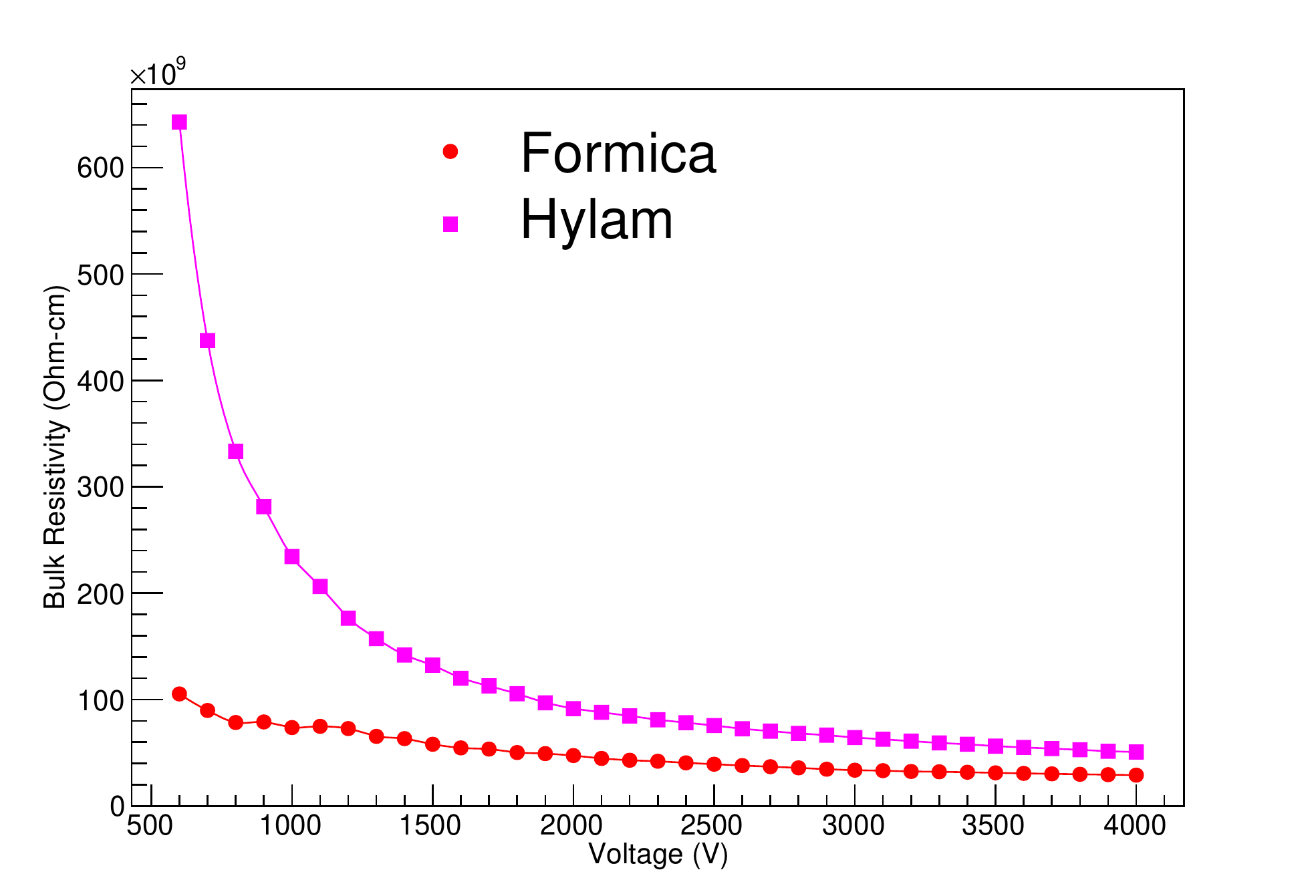}
\caption{\label{bpic} The bulk resistivity for Formica is stable throughout the applied range of voltages with respect to Hylam bakelite.}
\end{figure}

For Formica and Hylam, the measured values of bulk resistivity are 2.7 x 10$^{10} \Omega$ cm and 4.7 x 10$^{10} \Omega$ cm respectively at relative humidity of 44\% and temperature at 21$^{\circ}C$. The measured bulk resistivity of Formica and Hylam bakelite as a function of voltage is shown in Figure~\ref{bpic}. At a given applied voltage, the reproducibility in the measurement can vary within few percent due to the residual effect of the material polarization. These resistivity values are high enough to use these materials effectively as RPC electrodes. 

\subsection{\it Surface resistivity}\label{surface} 
\begin{figure}
\centering
\includegraphics[height=5cm,width=5cm]{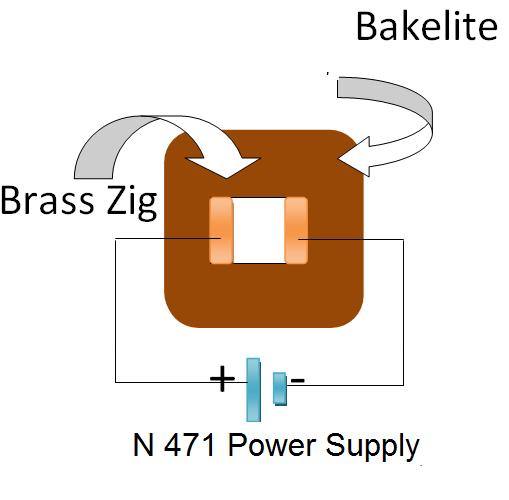}
\caption{\label{sset} The set-up for surface resistivity measurement.}
\end{figure}

The surface resistivity is measured on inner surfaces of both the bakelite samples (silicon coated or uncoated) before fabricating the detectors, which gives the indication about the unevenness of the surface texture. The surface resistivity of the bakelite sheet is measured using the set-up shown in  Figure~\ref{sset}. Two brass bars of soft padded conducting edges at the bottom have been taken for this experiment, which are placed on the bakelite sample. The bars, forming the opposite sides of a square shape, are mounted on epoxy plates. The length of brass bars as well as their separation is kept at 3 cm. A DC bias voltage is applied on the jig and leakage current through the sample (across the terminals of jig) is measured.
 
\begin{figure}[tbp]
\centering
\resizebox{6.5cm}{5.0cm}{\includegraphics[height=4cm,width=7cm]{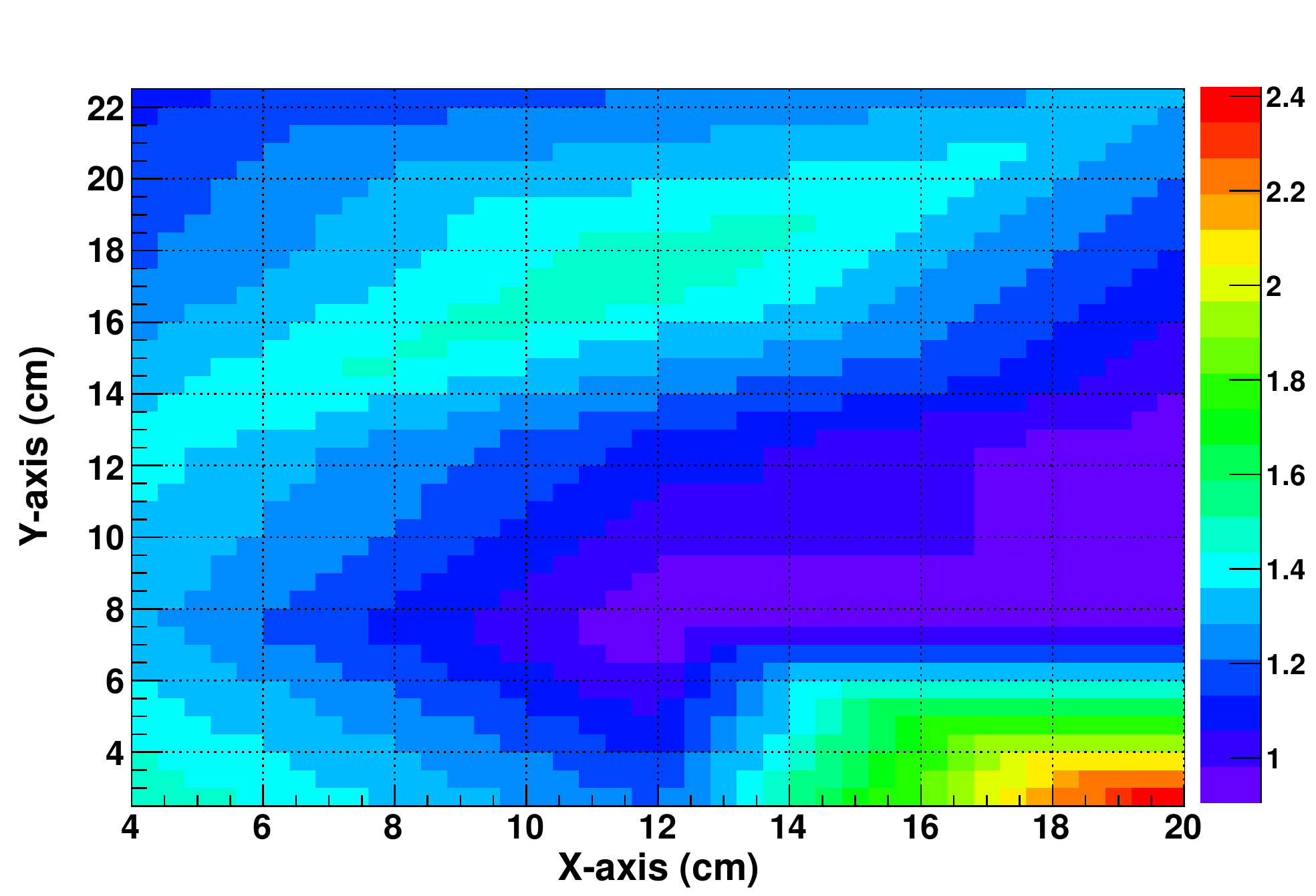}}
\resizebox{6.0cm}{5.0cm}{\includegraphics[height=5cm,width=7cm]{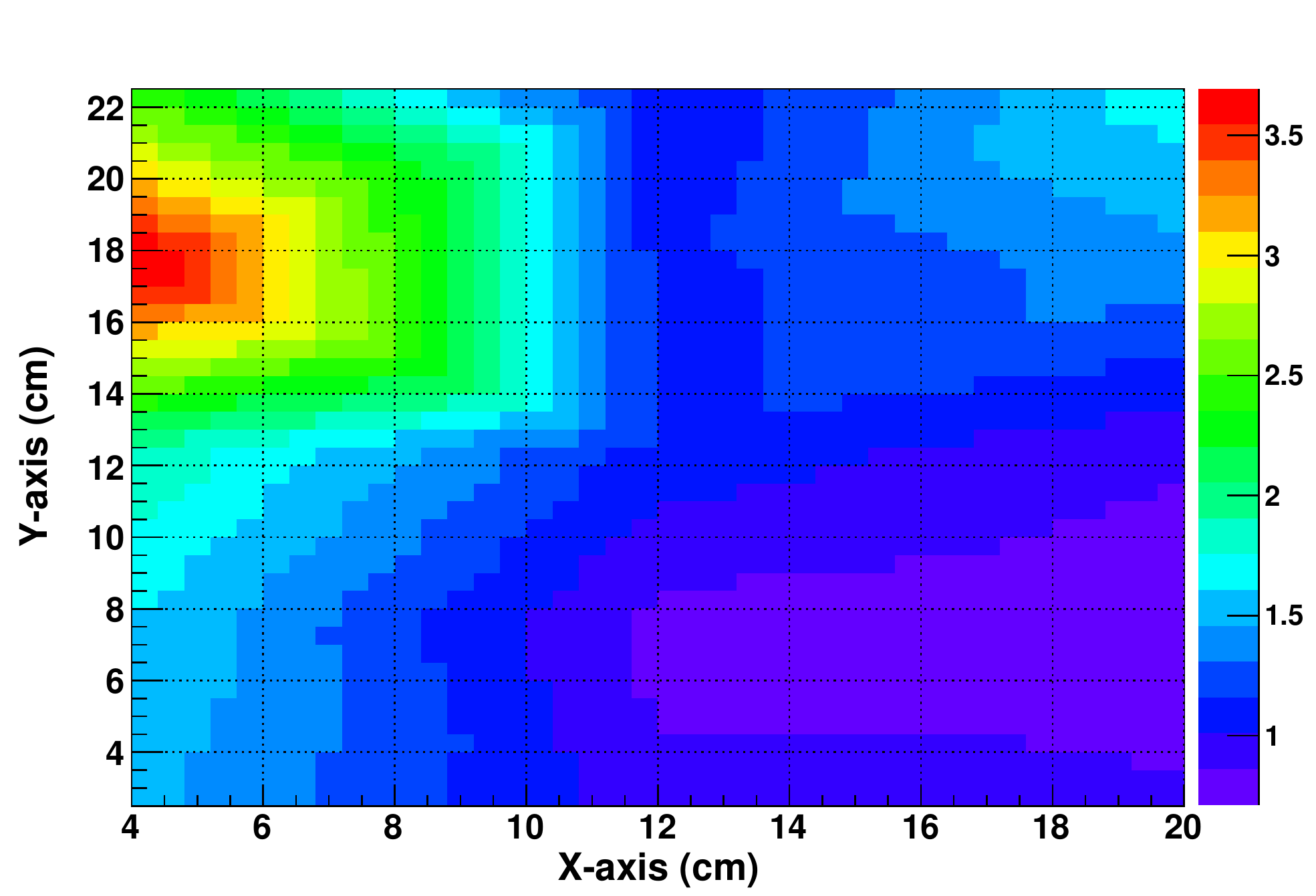}}
\caption{\label{spic} The surface resistivity ($10^9\Omega/\square$) contours for Formica (left) and Hylam (right) bakelites.}
\end{figure}

 Figure~\ref{spic} shows the surface resistivity variation in 2-dimensional space for both the samples at relative humidity of 42\% and temperature at 21$^{\circ}C$. The surface resistivity of the silicon coated (unpolymerised) surface is found to be less as compared to the one without silicon coating. 

\section{Detector assembly, gas mixing and cosmic test} \label{asmbl}
The RPCs were fabricated after exploring the structural and electrical properties of the electrodes. The gas is mixed in the desired proportion and mixture is then  flushed into the chambers. The chambers are tested using cosmic hodoscope. The details of RPC fabrication and testing is summarized below:

\subsection {\it Chamber fabrication}\label{fabrication}
RPC prototypes of 30 cm $\times$ 30 cm in sizes are fabricated. A thin layer of silicon coating is applied on inner surface of the electrode for better surface texture. The gap between two electrodes is kept small, 2 mm, for better time resolution of the RPC detector \cite{mihee}. The uniformity in the separation is maintained using polycarbonate spacers. Two nozzles, one for gas inlet and one for outlet, are used. The gap was filled with suitable gas mixture. The whole detector is sealed pack by using adhesive at the corners and side spacers. The RPC signals are picked by the copper strips, placed above and below the gas chamber for 2-dimensional readout. 
  
\subsection{\it Gas mixing}\label{gasmixing}
The proper gas mixing plays a vital role in detector response in detecting charged particles. For this propose, a gas mixing system capable of mixing and distributing gases is used. The constituent gases are mixed in the required proportion into a buffer cylinder and distributed into the RPC. The mixture of three gases is prepared on the basis of their properties and functionality. An electronegative gas such as Freon is used as the active medium gas. Isobutane gas was used as a quenching medium and absorbs ultraviolet photons produced in the electron-ion recombination. Sulfur Hexafluoride (SF$_{6}$) is used for controlling the production of extra electrons, if any. The enclosed gas mixture within chamber permits the detection of charged particles after applying high electric field. 

\subsection{\it Efficiency measurement}\label{cosmic}
The scintillator detectors are chosen as triggering detectors for cosmic (muon) signal. They also make full coincidence system for studying efficiency of RPC detector. Three scintillator paddles were arranged above and below the RPC detector. The coincidence counts from three scintillation detectors defines the 3-fold counts. Consequently, counts from coincidence with interleaved RPC detector defines the 4-fold counts. The RPC efficiency has been measured as a ratio of 4-fold counts to the 3-fold counts.

\section{RPC performance with different gas compositions, temperature and humidity}\label{results}
\begin{figure}[tbp]
\centering
\resizebox{7.5cm}{7cm}{\includegraphics{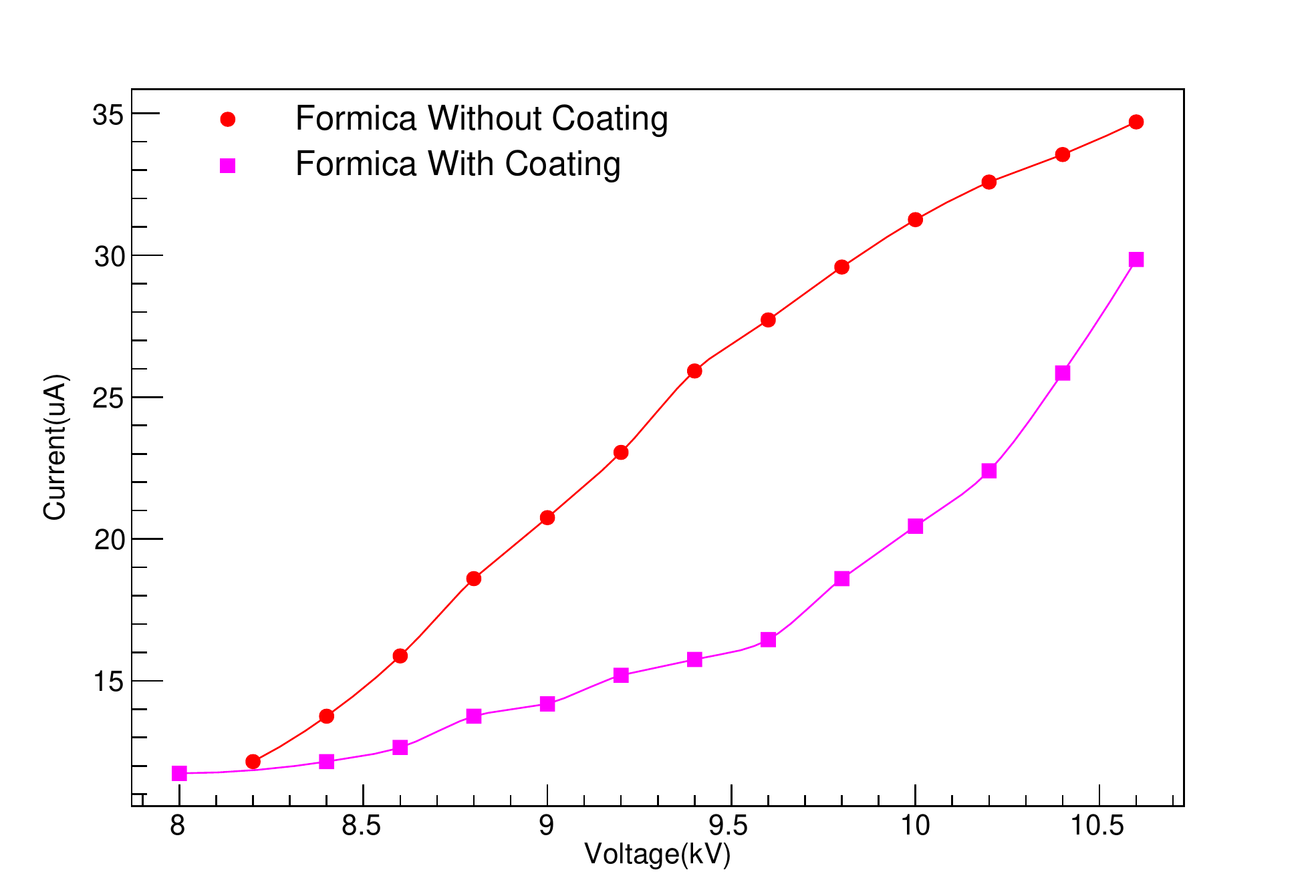}}
\resizebox{7.5cm}{7cm}{\includegraphics{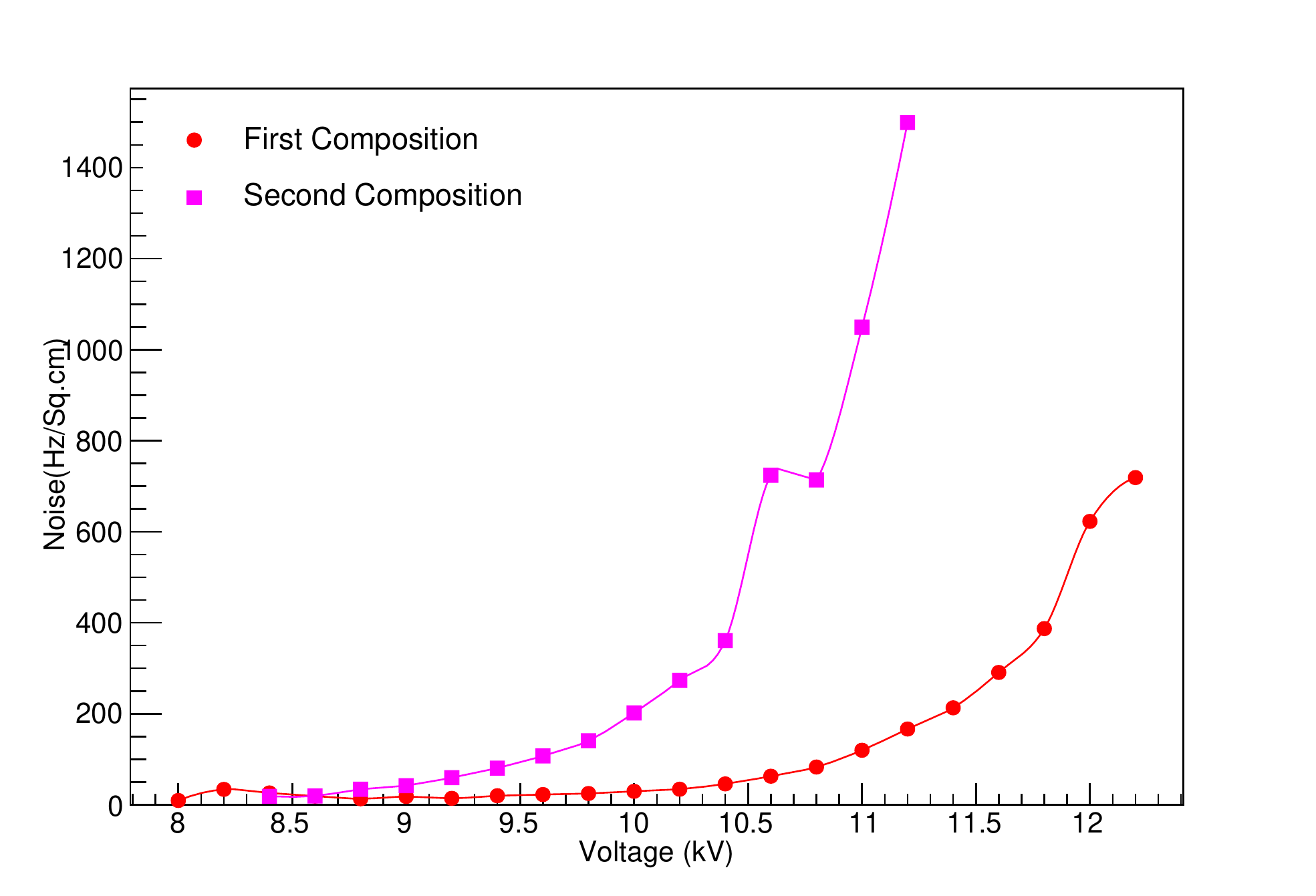}}
\caption{\label{eff1} The leakage current is less for silicon coated bakelite RPC (left). In the case of Formica bakelite, there is significant pull towards better stability for the noise rate under better gas composition (right).} 
\end{figure}

The electrode resistivity mainly determines the rate capability, chamber gap determines timing properties and gas mixture decides charge production. For the chamber fabricated from silicon coated Formica bakelite, the leakage current is less than the one without coated surface as shown in Figure~\ref{eff1}. The noise rate is also observed to be less for silicon coated bakelite RPC. This is expected as poor surface uniformity gives poor response. Comparable value of leakage current for Hylam based bakelite is observed, but it suddenly achieves breakdown after crossing threshold. The Bakelite samples have been procured from local market in the city of Delhi. The relative humidity in Delhi varies from 40\% to 90\% in the summer and rainy session. So there is a possibility that these samples might have been kept for long in the local market and hence have been exposed to very high humidity conditions. The efficiency of Formica based RPC detector comes to be around 90$\%$. This study has been extended under two different gas compositions namely CH$_{2}$FCF$_{3}$/C$_{4}$H$_{10}$/SF$_{6}$ = 67.7/32/0.3 (i), 95/4.5/0.5 (ii). A significant pull have been found towards better stability under first gas composition for the noise rate as shown in Figure~\ref{eff1} (right) . 

\begin{figure}[tbp]
\centering
\resizebox{7.5cm}{7cm}{\includegraphics{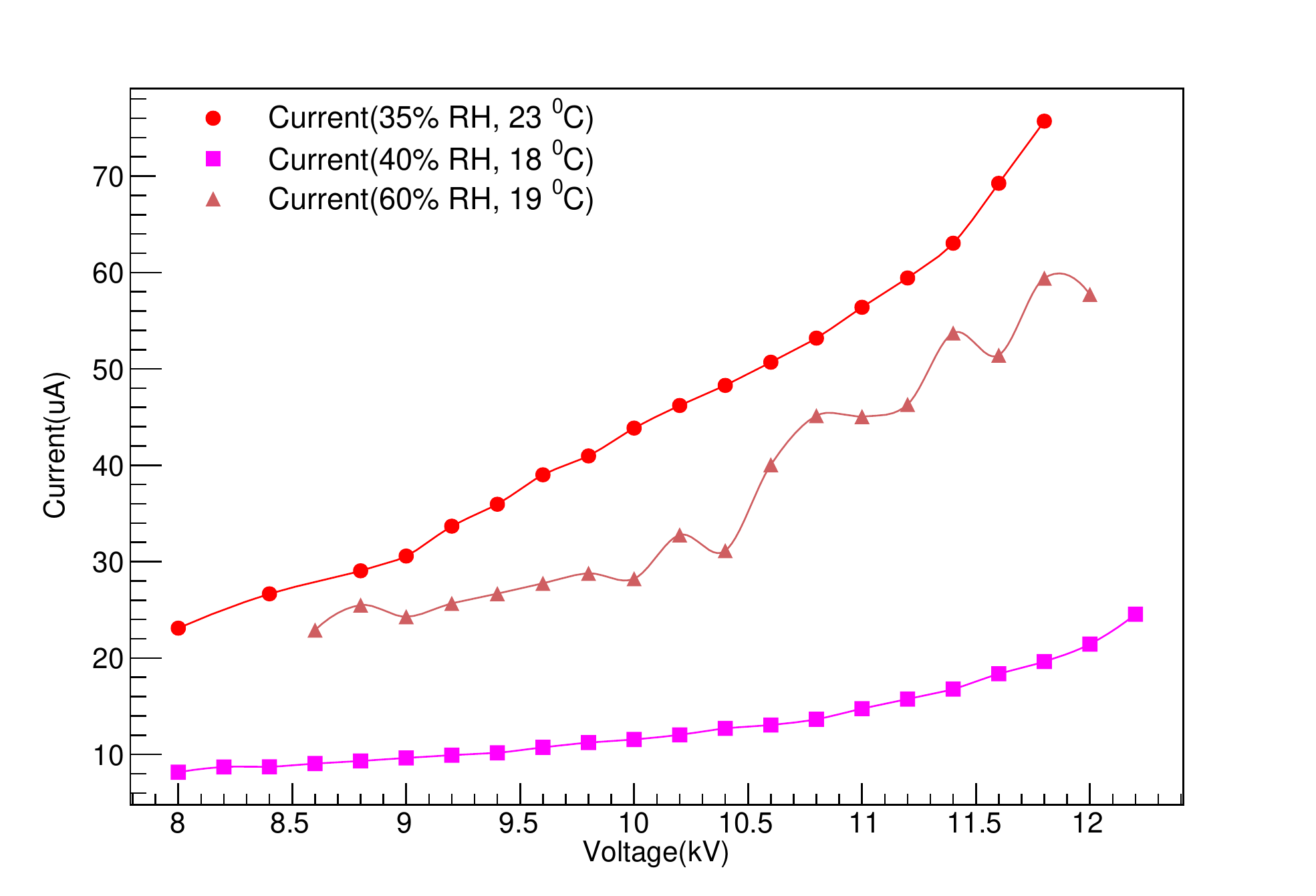}}
\resizebox{7.5cm}{7cm}{\includegraphics{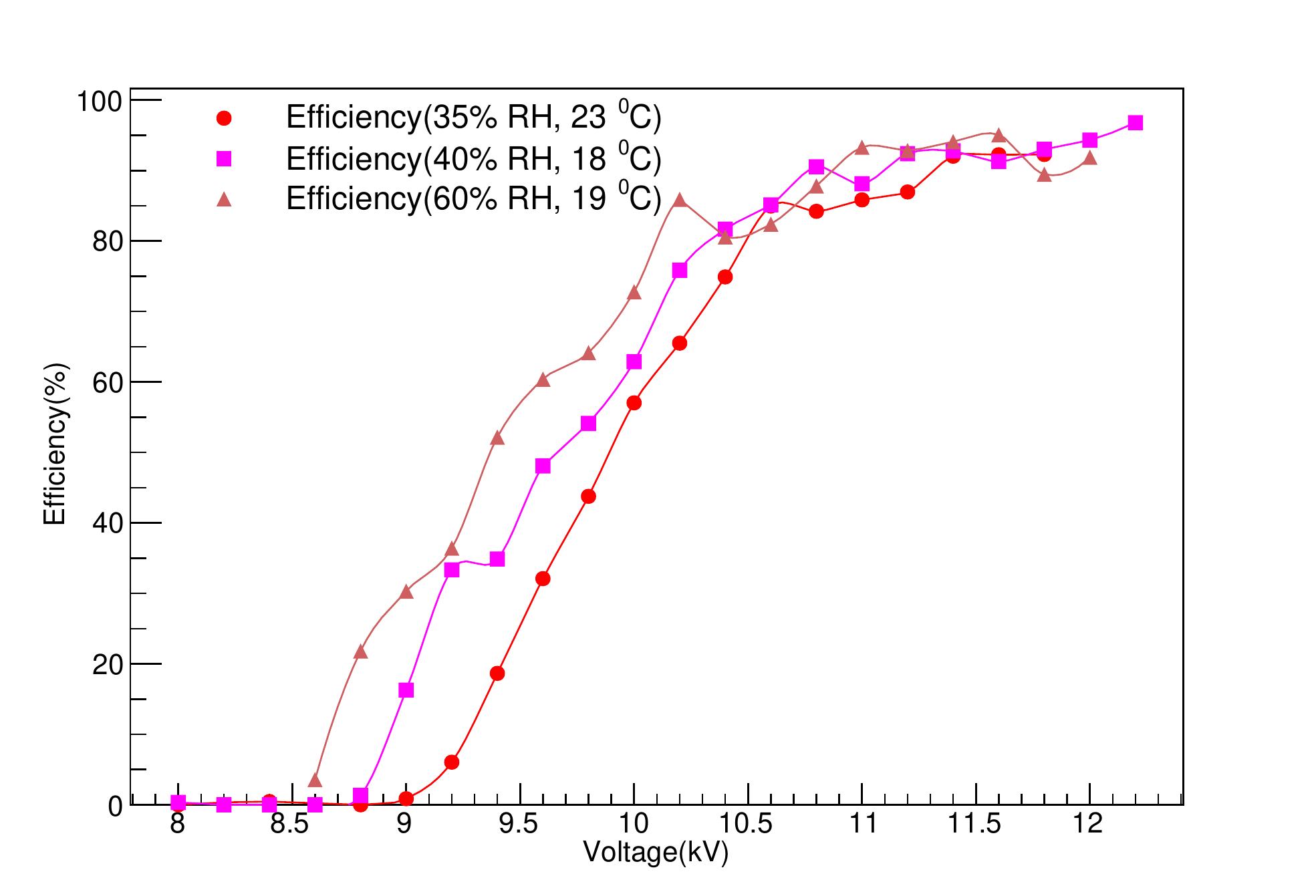}}
\scriptsize{\caption{ \label{current} In the case of Formika bakelite, the leakage current increases with the increase in temperature and also fluctuate with the increase in relative humidity (left) and threshold in the efficiency curve slightly changes but not the plateau length (right). }}
\end{figure}

 The performance of Formica RPCs with more stable (second) gas composition under different room temperature and relative humidity (RH) conditions, such as, (i) 23$^{\circ}C$, 35$\%$ (ii) 18$^{\circ}C$, 40$\%$ and (iii) 19$^{\circ}C$, 60$\%$ have been studied. Figure~\ref{current} shows the effect of variations in temperature and relative humidity on the performance of Formica  RPC. The leakage current has slight fluctuations at higher relative humidity as shown in Figure~\ref{current}. The length of the plateau in efficiency curve remains same, though threshold shifts.

\section{Results}
The R$\&$D effort for the bakelite RPC is presented keeping in view their use in upcoming India based Neutrino observatory (INO). The INO need $\sim$ 30,000 RPCs of 2 m $\times$ 2 m size, so the need is essentially to have low cost and readily available solutions. Bakelite RPCs of  30 cm $\times$ 30 cm were fabricated from different electrode materials. The characterization analysis of various types of electrode material have been performed for the selection of best RPC electrodes. Since Hylam and Formica are available locally, these have been chosen for our studies. Bakelite Formica shows better results as compared to Hylam. Although many factors can, in principle, influence the performance of the RPC, the effect of the nature of RPC electrode material have been considered in detail. The noise rate also increases with the increase in the temperature and relative humidity. For the detector kept under slightly different relative humidity conditions, little difference in the detector efficiency was observed. The reason for low efficiency (90$\%$) is misalignment of triggering detector. With revised trigger detector settings efficiency was found to be more than 95$\%$. 

\section{Conclusions}
The properties of the bakelite material affects the RPC performance. Since silicon coating improves the surface texture of electrodes, decrease in the RPC noise rate and leakage current have been observed. The electrodes having different resistivity values carry different RPC rate capabilities. The bakelite resistivity strongly depends upon the ambient conditions. The deterioration in the detector characteristics, such as noise rate and leakage current, have been observed. This is confirmed via a study performed on RPCs fabricated from two locally available bakelite samples and kept under different environmental conditions. 

\acknowledgments

We would like to acknowledge the financial support received from Department of Science and Technology (DST). Daljeet Kaur would like to thank Council of Scientific and Industrial Research (CSIR), India for the financial support. We would like to thank INO group of TIFR, Mumbai for providing some raw materials. We also thank Prof. J. P. Singh of IIT Delhi for helping us in carrying AFM measurements in his laboratory.


\begin{thebibliography}{9}

\bibitem{santonico}
 R.Santonico and R.Cardarelli, \emph{Development of resistive plate coumters},
 \emph{NIM} \textbf{A 187} (1981) 377.

\bibitem{pramana}
 S. Bhide et al., \emph{Preliminary results from India - based Neutrino Observatory detector $R\&D$ programme}, \emph{Journal of Physics}, \textbf{69(6)} (2007) 1015-1023.

\bibitem{atmnu}
S. Atthar et al., \emph{Technical Report of INO},
\emph{2006, vol I}.

\bibitem{meghna}
K.K Meghna et al., \emph{Measurement of electrical properties of electrode materials for the bakelite Resistive Plate Chambers},
\emph{2012 JINST 7 P 10003/1748-0221}.

\bibitem{dc1}
C. Lu,
\emph{RPC electrode material study},
\emph{NIM} \textbf{A 602} (2009) 761.

\bibitem{dc2}
S. Biswas et al., \emph{Performances of silicone coated high resistive bakelite RPC},
\emph{NIM} \textbf{A 661} (2012) S94-S97.

\bibitem{imppar}
G. Aielli et al., \emph{Electrical conduction properties of phenolic-melaminic laminates},
\emph{NIM} \textbf{A 533} (2004) 86.

\bibitem{mihee}
Mihee Jo et al., \emph{Gas Mixture Dependence of the Performance for Multigap Resistive Plate Chambers in the Avalanche Mode},
\emph{Journal of the Korean Physical Society} \textbf{56(5)} (2010) 1423-1429.

 
\end{thebibliography}
\end{document}